\documentclass{article}

\usepackage{arxiv}

\usepackage[utf8]{inputenc} % allow utf-8 input
\usepackage[T1]{fontenc}    % use 8-bit T1 fonts
\usepackage{hyperref}       % hyperlinks
\usepackage{url}            % simple URL typesetting
\usepackage{booktabs}       % professional-quality tables
\usepackage{amsfonts}       % blackboard math symbols
\usepackage{nicefrac}       % compact symbols for 1/2, etc.
\usepackage{microtype}      % microtypography
\usepackage{lipsum}
\usepackage{graphicx}
\usepackage{amsmath}
\graphicspath{ {./images/} }

\providecommand{\e}[1]{\ensuremath{\times 10^{#1}}} % uso 3.4\e{-3}

\title{Features of Linear Models that May Compromise Model-Based, Plant-Wide Control Techniques. The Case of the Tennessee Eastman Plant.}

\author{
 Sergio F. Yapur \\
  Facultad de Ingenier\'ia Qu\'imica \\
  Universidad Nacional del Litoral\\
  Santiago del Estero 2829 (3000) Santa Fe \\
  \texttt{syapur@fiq.unl.edu.ar} \\
}

\begin{document}
\maketitle
\begin{abstract}
This work examines a set of features that impact the reliability of linear models within the context of plant-wide control design (PWC). The study case is the Tennessee-Eastman (TE) plant. This benchmark problem is well-known for challenging many control design approaches. Analyses involve eigenvalues, average errors between simulations, condition numbers, and loss of rank across frequencies. These studies offer guidance for designing an effective plant-wide control system based on linear models.
\end{abstract}

% keywords can be removed

\section{Introduction}

Linear models present remarkable advantages in control system design, rendering a wide acceptance in the community. Some of these benefits come from the results of the long-established theory of linear dynamical systems, such as complete characterization of trajectory types, easy stability assessment, and a vast amount of methods for solving linear systems. In particular, linear Time-Invariant (LTI) systems have been a cornerstone for most modern control theory developments. 

On the other hand, most real-world models exhibit non-linearity to some extent. A linearization method, usually based on Taylor-type approximations, bridges the two realms. The method selected for this step is essential, although generally overlooked. A reason for this might be overconfidence built over the basis of a great many successful applications of Taylor approximations, mainly for simple systems.

Nonetheless, the situation may change whenever the system complexity is high enough. This kind of system arises naturally in the field of plant-wide control. The complexity of the problem may increase the computational cost of the linearization method. At the same time, it may degrade the reliability of the resulting linear model. In extreme cases, such linearization might severely mislead the control system design. This work examines some of the conditions for this to happen.

This article focuses on the Tennessee Eastman (TE) plant, a recognized benchmark problem within the PWC research community. This problem shares many typical characteristics with other complete chemical facilities, such as multi-variable inputs and outputs, open-loop instability, non-linearity, different sampling rates across signals, strong interactions between variables, and operative restrictions on process conditions. 

The original article of the TE plant mentions a linearization obtained from the nonlinear plant, together with some basic outcomes. The following paragraphs reproduce and expand those results. 

The first section of this article introduces the main features of the plant under study. Then, Section 2 overviews some basic observations that will guide the rest of the text. It also introduces the linearization codes under study. The following section develops comparative analyses that may impact model-based control system design from a numerical standpoint. The last part presents conclusions and recommendations. 

\section{The Tennessee Eastman Plant}

This test problem from the Eastman Company arose to supply the demand of the PWC research community \cite{DownsVogel} in the early 1990s. Despite the age of this problem, it is still valuable because of its complexity and realistic features. 

The problem is completely based on an actual process, although the company modified some details to protect proprietary rights. Essentially, the plant consists of a chemical reactor and a separation stage. The reactor produces the objective compounds, while the separation stage extracts these compounds from the rest of the reagents. Additionally, the unreacted reagents flow back to the reactor to increase profitability.

\begin{figure}[ht]
  \centering
  \makebox[\textwidth][c]{\includegraphics[width=0.8\textwidth]{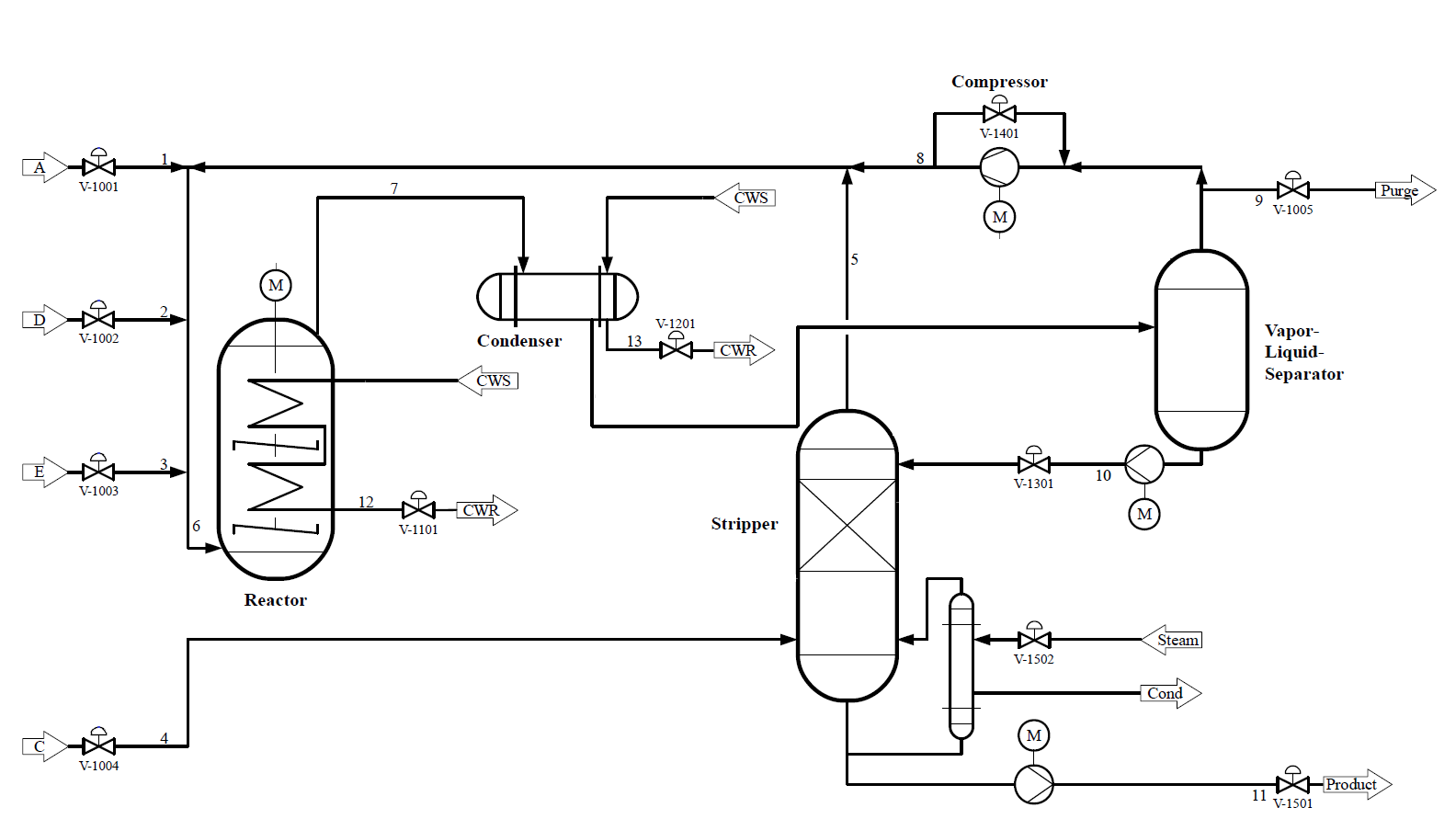}}
  \caption{Flow sheet of the TE problem. }
  \label{fig:Lorenz}
\end{figure}

The actual representation of the plant is an algorithm, originally written in Fortran and later reviewed and translated to C++ and Matlab\textsuperscript{\textregistered} \cite{Ricker_Archive}. The implementation used in this work is a Simulink\textsuperscript{\textregistered} block linked to an S-Function. No mathematical expressions are available, so handling input-output data numerically is the only alternative available. This situation narrows to some degree the number of available linearization methods that might serve as a first-hand analysis. In other words, this feature excludes methods that rely on a specific nonlinear model structure, such as ARX-type models. Naturally, it also segregates traditional analytical techniques. % So commands like linapp are unsuitable as it uses ARX model

Gaussian noise is present in every output signal to make the simulation results more realistic. The standard deviation used follows typical values for the measurement type (e.g., temperature, pressure, or flow rate). The noise features sensing issues and transmission faults present in any industrial setup. 

The input-output behavior of the system is highly nonlinear; in fact, it is chaotic \cite{yapur2021controlledvariable}. The chaoticity condition is possibly due to the simultaneous exothermic reactions, the interactions between units (both downstream and upstream because of the recycle stream), and the mass and heat transfer phenomena within the stripper. A consequence of the chaotic behavior is dynamic instability, worsened by the noise in output signals.

Another feature of interest is the presence of hard constraints on the process. That is, at any time a selected variable surpasses its allowed operative range, the simulation stops immediately. This fact resembles the plant shutdown triggered by automatic safety systems. 

The plant is a MIMO system with 40 inputs and 41 outputs. Twelve of the inputs are the manipulated variables, while the rest are binary signals that enable preprogrammed perturbations at any desired time.

In addition, there are two types of output signals, continuous and discrete. The continuous signals include temperature, pressure, and flow rate measurements. They are so-called continuous because the sampling time can be set to any desired value, at least in principle. Besides, the discrete signals derive from slow process measurements, all related to chemical compositions. Two sampling times exist for these signals, with $T_{s1}=0.1$ h and $T_{s2}=0.25$ h. The existence of different sampling times is common-ground in real chemical facilities. A natural consequence from a linearization standpoint is a higher system complexity since the selected technique has to deal with mixed sampling times across different signals.

\section{Notes on Linearization}
\label{sec:linearization}

The bibliography on linearization techniques is vast. Hence, the most common results are ubiquitous and well-documented \cite{Stephanopoulos, Sontag}. As the subject is mature, this work will not review the underlying theory. Instead, the focus is on testing traditional linearization techniques against the TE plant.

It is worth noticing that the current work differentiates between linearization techniques, which are inherently attached to a particular operative point, and general identification methods based on linear models, where a given point in the controlled-variable space may not be relevant whatsoever. Even when the control systems community has widely adopted both approaches, only the former will be considered here.

Linear models are often considered mathematical simplifications of real-life systems, mainly nonlinear. It is also supposed that these simplifications always retain the dynamical behavior around the linearization point, as the theory suggests. These beliefs, together with the fact that they are relatively simple to handle computationally, set linear modeling as the basis of many control techniques. Nonetheless, this hegemonic acceptance has its drawbacks. For instance, a validation step should be performed systematically for the resulting linear model. However, because of the underlying belief in the reliability of linear models, this step is frequently skipped. In particular, having a reasonable agreement between the dynamical behavior of the nonlinear system and its linear counterpart is not guaranteed under some circumstances.

Fundamentally, problems may arise whenever an application of a given theorem lays outside the scope of the corresponding hypothesis. This rather obvious mistake in the academic community might be less evident in engineering practice, where the rush for getting results as soon as possible is nearly omnipresent. An almost trivial, nonetheless important example is the classic Taylor theorem, where functions shall be differentiable. Yet it is not rare to find Taylor-type linearizations over functions that handle digital signals. Therefore, the presence of discontinuities cast a mismatch between theory and practice. Strictly speaking, linearization techniques suppose continuous-time systems \cite{Sontag}. Nonetheless, adaptations to discrete-time systems are commonplace \cite{Klee}.

Moreover, several computational tools fail to detail the theoretical scope of applicability of the available functions, nor do proper bibliography references appear in the documentation. Consequently, the validation step becomes a research task on its own, draining a significant amount of time and resources.

\subsection{Linearization Codes}

The present article will not review every linearization technique available in the literature. Instead, it is limited to assessing the effectiveness of some standard linearization codes for the TE plant. Given the fact that the object under study is a Simulink\textsuperscript{\textregistered} model, any linearization method selected must keep compatibility with either this object class or related classes. Thus, this work utilizes the \texttt{linmod} family of Matlab\textsuperscript{\textregistered} commands for the sake of convenience. Additionally, as the plant uses mixed sampling-time signals, two approaches are compared. On one side, a continuous-time model that derives from the \texttt{linmodv5} code. On the other side, a discrete-time linear model that follows from the \texttt{dlinmod} algorithm. These models will be hereafter named L1TE and L2TE respectively. Both are state-space models.  

\section{Discussion and Results}

\subsection{Open-loop Eigenvalues} 

A complete characterization of the dynamics of a linear system is possible from the eigenvalues and eigenvectors of the matrix representation of the model, as known from the spectral decomposition theorem. A remarkable advantage of this analysis is revealing open-loop stability issues readily. Because of these applications, open-loop eigenvalues have become the first line of characterization of dynamical models. 

The following sections present eigenvalues for each linearization under revision. For each model, the operative point comes from the Simulink\textsuperscript{\textregistered} block. In turn, the block calls an initialization routine written in C++ \cite{Ricker_Archive}. Ultimately, this means that the operative point is the same as the one used in the original article \cite{DownsVogel}. Furthermore, the initial state is equal to the operative point.

\subsubsection{L1TE Model}

Table \ref{tab:tablaAutovalorL1TE} shows the eigenvalues for L1TE, obtained from a standard linearization around the operative point of the plant. The table also includes multiplicity information. Given that eigenvalues $e_{17}, e_{18}, e_{25}, e_{26}$ are on the Right-Hand Side (RHS) plane, the instability of the original model remains on its linear counterpart L1TE. Additionally, the presence of poles near the origin confirms integral behavior, probably due to the vessels of the process units. Notice also the spread in magnitudes thorough the table. This spread means the model L1TE can reproduce both fast and low dynamics. Although this feature also increases the stiffness of the model. 

\begin{table}[ht]
	\begin{center}
		\caption{Open-loop eigenvalues for L1TE.}
		\label{tab:tablaAutovalorL1TE}
		\resizebox{16.0cm}{!}{
		\begin{tabular}{|ccc||ccc||ccc|} 
			\hline 
			\textbf{Symbol}   & \textbf{Value} & \textbf{Mult.} &
			\textbf{Symbol}   & \textbf{Value} & \textbf{Mult.} & 
			\textbf{Symbol}   & \textbf{Value} & \textbf{Mult.} \\ 
			\hline
			$e_1$           &  -1968.1              & 1 &      
			$e_{14}$        & -26.228               & 1 &      
			$e_{28}$        & -0.081437             & 1  \\
			$e_2$           & -726.37               & 1 &
			$e_{15},e_{16}$ & -11.619$\pm$4.4903i & 1 & 
			$e_{29}$        & -0.074422             & 1 \\
			$e_3$           & -529.7                & 1 &      
			$e_{17},e_{18}$ & 3.0648$\pm$5.0837i    & 1  & 
			$e_{30}$        & -9.3925\e{-7}         & 1  \\ 
			$e_4$           & -139.16               & 1 &      
			$e_{19}$        & -13.116               & 1 &     
			$e_{31}$        & -1.2122\e{-10}        & 1  \\
			$e_5$           & -114.19               & 1 &      
			$e_{20}$        & -12.832               & 1 &     
			$e_{32}$        & -5.1784               & 7   \\
			$e_6$           & -87.025               & 1 &      
			$e_{21}$        & -10.76                & 1  &      
			$e_{33}$        & -450                  & 2 \\
			$e_7$           & -81.858               & 1 &      
			$e_{22}$        & -9.9354               & 1 &     
			$e_{34}$        & -600                  & 1 \\
			$e_8,e_9$       & -37.22$\pm$17.18i   & 1 & 
			$e_{23}$        & -8.9211               & 1 & 
			$e_{35}$        & -400                  & 1 \\
			$e_{10},e_{11}$ & -34.797$\pm$14.934i & 1 &            
			$e_{24}$        & -1.2219               & 1 & 
			$e_{36}$        & -514.29               & 1 \\
			$e_{12}$        & -29.275               & 1 &  
			$e_{25},e_{26}$ & 0.024974$\pm$0.15521i & 1 & 
			$e_{37}$        & -720                  & 6 \\
			$e_{13}$        & -28.835               & 1 &  
			$e_{27}$        & -0.11529              & 1 &    
			$e_{38}$        & -30                   & 1  \\
			\hline
		\end{tabular}
		}
	\end{center}
\end{table}

A remarkable feature of Table \ref{tab:tablaAutovalorL1TE} is that the largest negative eigenvalue $e_1$ agrees with the one declared in the original work \cite{DownsVogel}. This fact constitutes a partial validation of the linear model, although the authors did not provide any additional eigenvalues to compare.

\subsubsection{L2TE Model}

Recalling that L2TE is a discrete-time model, it has an additional parameter $T_s$, the sampling time. This parameter is set to $T_s=2.5\e{-4}$ h, corresponding with twice the maximum frequency observed in L1TE. This sampling time lets L2TE retain the same dynamic range as the L1TE, in the sense of the Nyquist sampling theorem. In other words, the L2TE model is capable of reproducing fast dynamics as well as the L1TE model.

Table \ref{tab:tablaAutovalorL2TE} present the eigenvalues for L2TE. Because the L2TE is a discrete model, the interpretation from the eigenvalue locus differs from the continuous case. Similar to the analysis of the L1TE eigenvalues, an integral behavior occurs through $e_{30}$, and instabilities appear through the conjugate pair $e_{31},e_{32}$ since they locate off the unit circle. However, this model does not present any oscillatory behavior, as there are no values on the Left-Hand Plane (LHP). This difference regarding the L1TE model has profound consequences since no resonance phenomena are possible.

\begin{table}[ht]
	\begin{center}
		\caption{Open-loop eigenvalues for L2TE.}
		\label{tab:tablaAutovalorL2TE}
		\resizebox{16.0cm}{!}{
		\begin{tabular}{|ccc||ccc||ccc|} 
			\hline 
			\textbf{Symbol}   & \textbf{Value} & \textbf{Mult.} &
			\textbf{Symbol}   & \textbf{Value} & \textbf{Mult.} & 
			\textbf{Symbol}   & \textbf{Value} & \textbf{Mult.} \\ 
			\hline     % comandos utiles: \e{-7}, $\pm$
			$e_1$           &        0.6114             & 1 &     
			$e_{11}$        &        0.9785             & 1 &
			$e_{23},e_{24}$ &        0.9971$\pm$0.0021i & 1 \\
			$e_2$           &        0.8339             & 1 &      
			$e_{12}$        &        0.9797             & 1 &
			$e_{25}$        &        0.9973             & 1 \\
			$e_3$           &        0.8353             & 6 &
			$e_{13},e_{14}$ &        0.9907$\pm$0.0043i & 1 &
			$e_{26}$        &        0.9975             & 1 \\
			$e_4$           &        0.8607             & 1 &
			$e_{15},e_{16}$ &        0.9913$\pm$0.0037i & 1 &
			$e_{27}$        &        0.9978             & 1 \\
			$e_5$           &        0.8760             & 1 &
			$e_{17}$        &        0.9925             & 1 &
			$e_{28}$        &        0.9987             & 7 \\
			$e_6$           &        0.8794             & 1 &
			$e_{18}$        &        0.9927             & 1 &
			$e_{29}$        &        0.9997             & 1 \\
			$e_7$           &        0.8936             & 2 &      
			$e_{19}$        &        0.9928             & 1 &     
			$e_{30}$        &        1.0000             & 7 \\
			$e_{8}$         &        0.9048             & 1 &
		    $e_{20}$        &        0.9935             & 1 &
		    $e_{31},e_{32}$ &        1.0008$\pm$0.0013i & 1 \\
		    $e_{9}$         &        0.9658             & 1 &
		    $e_{21}$        &        0.9967             & 1 &
		                    &                           &   \\
		    $e_{10}$        &        0.9719             & 1 &
		    $e_{22}$        &        0.9968             & 1 &
		                    &                           &   \\
		    \hline
		\end{tabular}
		}
	\end{center}
\end{table}

From the eigenvalues of Table \ref{tab:tablaAutovalorL2TE}, it is clear that there are no matches regarding the linear model addressed in the original work \cite{DownsVogel}. Consequently, one can conclude that Downs and Vogel did not consider this particular kind of linearization.

\subsection{Deviations from Nonlinear Model}

\begin{figure}[ht]
  \centering
  \makebox[\textwidth][c]{\includegraphics[width=1.2\textwidth]{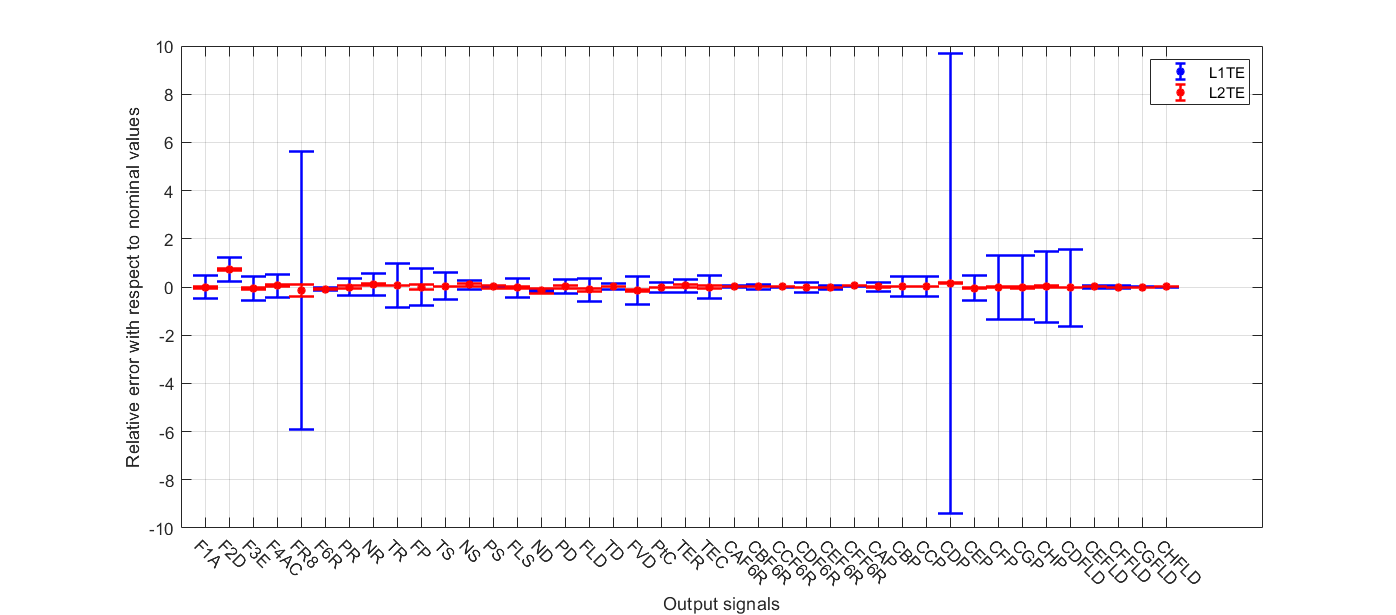}}
  \caption{Mean values of variable deviations with respect to nonlinear model simulation}
  \label{fig:CompLin}
\end{figure}

A good measure of the relative quality of linearization models is the average deviation from the original nonlinear system. Due to the instability of the open-loop nonlinear system, the simulation usually runs for up to an hour before reaching shutdown conditions. During this time, some variables are increasingly further from the operative point, leading to growing deviations. By comparing measures of these deviations between models, one may acquire a sense of linearization reliability. Fig. \ref{fig:CompLin} presents this comparison across every output variable using an error-bar plot. In the x-axis are acronyms to identify process variables. For this work purpose, detailed descriptions of every process variable are unnecessary. Anyhow, for general guidance, names starting with F refer to flow variables, N refers to level, T refers to temperature, P refers to pressure, and C refers to composition.

Because of the diverse nature of these variables, a side-to-side comparison on the same scale may be deceiving. This effect is evident when process units are different for two given variables. Nonetheless, different scales occur also when variables are of the same type. For instance, the nominal value for the flow rate F1A is around 0.25 [kscmh\footnote{Kilo standard cubic meter per hour}], while it is 42.33 [kscmh] for flow rate F6R. So relatively small deviations are more significant for F1A than for F6R. A single, normalized scale for all variables resolves this issue, defined as the relative deviation of the average value of the simulation concerning the nominal quantities:

\begin{equation}
    \label{eq:relValues}
    y_j = \frac{ \left( \overline{y_j^{nl}(t_i)} - y_j^{nom} \right) \cdot 100} {y_j^{nom}}.
\end{equation}

Notice that values $y_j$ refer only to the nonlinear model TE. Therefore, this set of values is independent of any linearization, making them suitable to assess modeling performance. Also, symbol $y_j^{nom}$ stands for nominal value, corresponding to the operative point \cite{DownsVogel}. In addition, the over-line denotes the mean value. Other measures of central tendency are possible. However, since the Gaussian distribution of the noise is symmetric, the mean value seems adequate because it compensates for noise errors to some extent.

Every computation $y_j$ from Eq. \ref{eq:relValues} is depicted in Fig. \ref{fig:CompLin} for a comprehensive comparison. The graphic represents these values with a dot placed at the center of each vertical interval. The spread of such intervals denote the relative errors of each linear model, according to the following expression: 

\begin{equation}
    \label{eq:relErrors}
    e_j = \frac{ \overline{y_j^l(t_i) - y_j^{nl}(t_i)}}{y_j^{nom} }. 
\end{equation}

Here, $y_j^l(t_i)$ and $y_j^{nl}(t_i)$ represent signal outputs from linear and nonlinear model, respectively. As in Eq. \ref{eq:relValues}, each deviation is divided by the corresponding nominal value to correct for different unit scales. Moreover, notice that output signals $y_j^{nl}$ might change from simulation to simulation due to the combined effect of chaos and noise in the TE model. However, in practice, minor differences are observed for simulations running over a time of one hour or less.

From Fig. \ref{fig:CompLin} it is clear that the L2TE model is substantially superior regarding deviations from the source model for nearly every output variable. In other words, the discrete linearization shows better general behavior regarding the nonlinear plant. This fact might have benefited many works based on the TE plant.

Finally, it is possible to summarize the relative deviations in a single indicator by summing them up. By doing this, the L1TE model gives out a value of 31.2 against 1.2 of the L2TE model. Overall, deviations from L1TE are 26 times larger than those of L2TE.

%\subsubsection{A review of TE linearization}

%Due to the existence of mixed-sampling across output-signals, one might think it is reasonable to group sets of input-output signals by output sampling rate, and then use adjusted linearization schemes for each set. For this approach to work properly, it is required ....
%A fundamental problem with this approach comes from a combination of factors. On the one side, random noise added to output signals would make each simulation unique regarding the input-output behaviour. On the other side, the chaotic nature of the plant would amplify any little deviation, even if it comes from the added noise. As result, it is extremely unlike that ...

\subsection{Condition number with respect to Inversion}

Several control design techniques rely on a matrix inversion, a pseudo inverse, or related operations. Centralized approaches are good examples of this assertion. A typical MIMO technique for designing controllers usually uses the inverse of the plant. Another example concerns decoupling strategies based on steady-state gain matrices. Moreover, several control system studies depend on numerous arithmetic operations. A few examples are the Niederlinski criterion, the Relative Gain Array \cite{Skogestad1996}, solving the Lyapunov equations, among many others. For all these cases, checking numerical stability is crucial for quality results.

The condition number $\gamma$ for inversion is defined as the quotient of the maximum singular value $\overline{\sigma}$ and the minimum singular value $\underline{\sigma}$ \cite{Skogestad1996}. With such a definition, the connection to the concept of singular values is straightforward. They measure the directional sensitivity of the input vector. Either a extremely high gain $\overline{\sigma}$ or low gain $\underline{\sigma}$ reveals system-level obstacles regarding achievable control quality \cite{Skogestad1996, Chin}.

In a numerical problem of type $\mathcal{A}x = b$, the condition number $\gamma$  measures the worst case of how much the output can change in proportion to small perturbations in the input data \cite{Faul}. A problem with a high condition number is said to be ill-conditioned. In control systems, matrix $\mathcal{A}$ can be regarded as the Transfer Function Matrix (TFM) $\mathbf{G}(s)$,

\begin{equation}
    \label{eq:TFM}
    \mathbf{Y}(s) = \mathbf{G}(s)\mathbf{U}(s).
\end{equation}

In this case, the condition number has a straightforward interpretation. Whenever $\gamma$ is high, a slight perturbation on the manipulated variables $\mathbf{U}(s)$ can largely deviate the desired output $\mathbf{Y}(s)$.

%Alternatively, $\mathcal{A}$ can also be considered as the state matrix $A$ in the state-space representation

%\begin{align}
%    \label{eq:SSequations}
%    \dot{x}(t) &= Ax(t) + Bu(t) \\
%         y(t)  &= Cx(t) + Du(t).
%\end{align}

%Here, it is possible to explicitly write the perturbation $\delta$ in terms of the matrices, $\delta = $

Besides the dynamic interpretation, a high condition number $\gamma$ generally indicates the presence of numerical issues, as it addresses a loss of precision through numerical manipulations. It is especially evident in matrix inversion, where a large enough $\gamma$ means that the matrix is close to being singular. However, numerical issues may appear beyond this particular operation. In principle, any vector manipulation may be potentially affected because the ultimate cause of the numerical error amplification is either extreme dilations or contractions. Although this fact is widely-known in numerical analysis, it is usually overlooked in control systems design, even when several controller-design techniques rely on matrix inversion.

\begin{figure}[ht]
  \centering
  \makebox[\textwidth][c]{\includegraphics[width=1\textwidth]{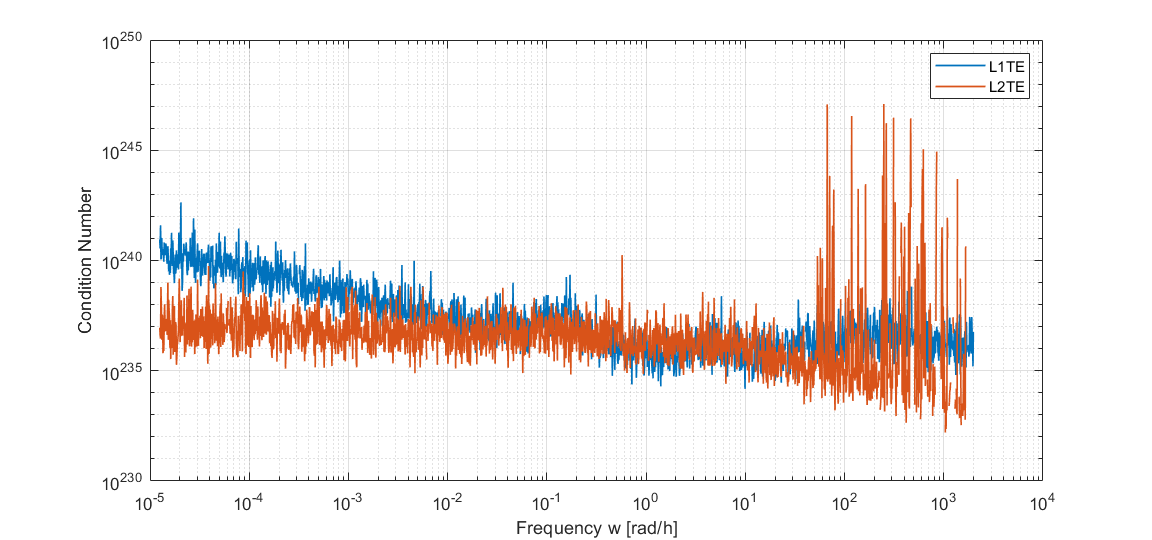}}
  \caption{Condition number against frequency for both linear models.}
  \label{fig:CompNC}
\end{figure}

The usual way to compute a condition number is at steady-state conditions. However, even when this calculation is formally possible, there is no actual steady-state in the open-loop nonlinear plant due to the severe instability of the system. For this reason, it may be more meaningful considering $\gamma$ across a range of frequencies of interest. These relations appear in Fig. \ref{fig:CompNC} for L1TE and L2TE linear models. Notice that both axes make use of a logarithmic scale. As evident from the figure, either linear model presents massive condition numbers. Perhaps the only relevant difference is that L1TE behaves slightly better at high frequencies, while L2TE is preferable at low frequencies. This difference may suggest that the L2TE model better captures low-frequency dynamics, possibly due to its discrete nature. In any case, these differences are relatively negligible compared to the absolute magnitudes of the numbers. 

The main conclusion from the above figure is that neither linearization is reliable from a numerical standpoint, especially regarding matrix inversions. As a result, several traditional analyses and design approaches from control theory either fail or underperform using models such as L1TE or L2TE. The numerical unreliability may be the root cause of difficulties while treating the TE plant. It was confirmed in a former work \cite{YapurTesisDI} that several control design methods indeed failed for linearization models. The ill-conditioning is so severe that even the computations for condition numbers themselves may differ from one computer system to another. 

For systems with these extreme condition numbers, two options may result in more promising: to explore another kind of linear modeling or to use nonlinear design techniques \cite{YapurTesisDI, Slotine}.

\subsection{Loss of Rank}

An alternative way to assess matrix inversion issues is through loss of rank. The rank of a TFM is the dimension of the subspace spanned by its rows. A well-known result states that the rank equals the number of nonzero singular values. From the previous study over the linear models, the magnitude of the high condition number is essentially due to an almost null minimum singular value $\underline{\sigma}$. Whenever a singular value decreases below the computer system $\epsilon$, a loss of rank may occur.

\begin{figure}[ht]
  \centering
  \makebox[\textwidth][c]{\includegraphics[width=1\textwidth]{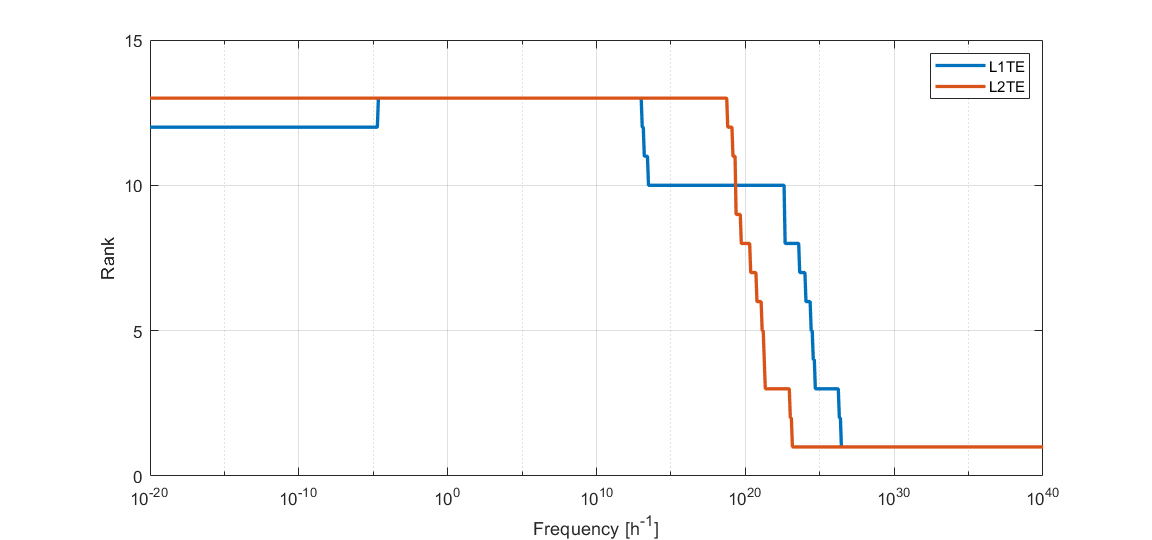}}
  \caption{Rank versus frequency for L1TE and L2TE models.}
  \label{fig:LossRank}
\end{figure} 

As was the case with condition numbers, studying the evolution of matrix rank across frequencies provides a better idea of the numerical stability of the system matrix. Figure \ref{fig:LossRank} present this analysis. Notice that if the frequency range is wide enough, several changes of rank values appear. Also, the maximum rank achievable is 13, a number significantly less than either of the MTF dimensions. Once more, this plot confirms the ill-conditioning of the system. Regarding each linear model performance, the L2TE is constant over a wider rank of frequencies than L1TE.

\section{Conclusions and Future Work}

Working with open-loop linear models requires special attention whenever complex systems are involved, as usual with PWC problems. Various factors can compromise the applicability of traditional linearization methods. For the case of the TE plant, one may enumerate these factors: chaotic behavior, low signal-to-noise ratio, mixed type of signals regarding continuity, absence of proper analytical expressions, among others. In general, any characteristic that excludes any underlying hypothesis of the intended linearization method may cast an unreliable model.

Regardless of the former difficulties, exploring linearization models may be helpful to expand the knowledge of the nonlinear system. For instance, the former analysis served to identify the spread of characteristic times when the TE plant is operating at a given operative point. It also delivered information on the system stability, a piece of valuable information in control system design. In any case, verifying the suitability of the linearization model to reproduce the nonlinear system behavior is strongly advisable.

Two models resulted from typical linearization techniques, one of them being continuous and the other discrete in time. The most significant difference between these models is that L2TE dynamic responses were significantly closer to the corresponding of the TE plant than its continuous-time counterpart L1TE. However, after analyzing condition numbers and loss of rank across frequencies, it was clear that both models are unreliable regarding numerical manipulations. Consequently, neither should be considered a solid basis for control system analysis and design. 

Another remarkable difference found between models comes from the eigenvalue analysis. It is worth recalling that both linearization models share the same operative point and reproduce the same frequency range dynamics. However, the L1TE model produces oscillatory output signals, while the L2TE does not. It is not a minor difference since, in particular, the discrete-time model lacks resonance phenomena.

Overall, it is advisable to assess potential numerical issues for any linearization model. Tools like condition number, singular values, average dynamical deviations, and rank stability were handy for this purpose. Nonetheless, further studies are possible, like input sensitivity or reproducibility of convergence. In particular, sweeping the signal-to-noise ratio in the input for both linear and nonlinear models may assist in detecting modeling errors.

Future works will explore alternative linear models for this plant, including modern identification methods. Their reliability compared to classical linearization models will be explored, as their usefulness in control system design.

%%%% HASTA ACA LLEGUE CON LA REVISION

\bibliographystyle{unsrt}

\end{document}